# Classification of COVID-19 in Chest CT Images using Convolutional Support Vector Machines

Umut Özkaya[1*], Şaban Öztürk[2], Serkan Budak[1], Farid Melgani[3], Kemal Polat[4]

[1] Dept. of Electrical and Electronics Engineering, Konya Technical University, Konya, Turkey
[2] Dept. of Electrical and Electronics Engineering, Amasya University, Amasya, Turkey
[3] Dept. of Information Engineering and Computer Science, University of Trento, Trento, Italy
[4] Dept. of Electrical and Electronics Engineering, Bolu Abant Izzet Baysal University, Bolu, Turkey
Corresponding Author: uozkaya@ktun.edu.tr

**Abstract:**

Purpose: Coronavirus 2019 (COVID-19), which emerged in Wuhan, China and affected the whole world, has cost the lives of thousands of people. Manual diagnosis is inefficient due to the rapid spread of this virus. For this reason, automatic COVID-19 detection studies are carried out with the support of artificial intelligence algorithms.

Methods: In this study, a deep learning model that detects COVID-19 cases with high performance is presented. The proposed method is defined as Convolutional Support Vector Machine (CSVM) and can automatically classify Computed Tomography (CT) images. Unlike the pre-trained Convolutional Neural Networks (CNN) trained with the transfer learning method, the CSVM model is trained as a scratch. To evaluate the performance of the CSVM method, the dataset is divided into two parts as training (%75) and testing (%25). The CSVM model consists of blocks containing three different numbers of SVM kernels.

Results: When the performance of pre-trained CNN networks and CSVM models is assessed, CSVM (7×7, 3×3, 1×1) model shows the highest performance with 94.03% ACC, 96.09% SEN, 92.01% SPE, 92.19% PRE, 94.10% F1-Score, 88.15% MCC and 88.07% Kappa metric values.

Conclusion: The proposed method is more effective than other methods. It has proven in experiments performed to be an inspiration for combating COVID and for future studies.

## 1. INTRODUCTION

A group of patients was infected with a novel coronavirus disease in Wuhan, China, in late December 2019. This virus was called "severe acute respiratory syndrome coronavirus 2" (SARS-CoV-2) before the world health organization (WHO) renamed it COVID-19 (Khan et al. 2020). After being infected, various symptoms such as fever, cough, weakness, and respiratory problems occur in patients. Depending on the state of the immune system and various personal factors, there are cases with pneumonia, multi-organ failure, and death (Mahase 2020). This virus, which has such severe symptoms, spreads so quickly that it took 30 days to spread from Wuhan to China (Wu and McGoogan 2020). In the United States, the total number of cases have exceeded 5.6 million since January 20, 2020, when it was first seen. The rate of spread is also high in other countries. As a result of the increase in the number of patients, patients with severe symptoms need to stay in hospitals. Some studies show that the average hospital stay is 22 days (Liu et al. 2020). Considering the condition and number of hospitals, it is very difficult to overcome this burden (Liew et al. 2020). For this reason, various drug and vaccine studies are still ongoing for definitive treatment (Rismanbaf 2020). Although there is no definitive treatment yet, pharmacological treatments are strongly needed in cases of severe disease (Scavone et al. 2020). Reverse Transcription-Polymerase Chain Reaction (RT-PCR) test is accepted as a standard for the definitive detection of this disease. However, in the early stages of the disease, this test fails in many cases (Shi et al. 2020a). For this reason, X-ray and Computed Tomography (CT) devices, which are easily available medical imaging devices, are used in clinical practice (Ozkaya et al. 2020). These imaging tools play a crucial role in the early diagnosis of COVID-19. Especially compared to test kits, they can produce faster results, analyze multiple cases at the same time, and give clear information about the progress of the case (Khan et al. 2020). With the examination of CT and X-Ray images, patients can be diagnosed, and the treatment process can be started and many lives can be saved. However, considering the large number of patients and the rate of spread of the disease, it is very difficult for experts to respond to this number. As seen in previous examples, manual examination by professionals in hospitals is very slow (Tanne et al. 2020). At this point, it is understood that intelligent technologies will make very important contributions to the diagnosis process. For this purpose, many researchers focus their attention on artificial intelligence (AI) studies that can accurately and quickly diagnose COVID-19 images.

AI methods contribute significantly to almost every field from the manufacturing industry to the healthcare industry (Jaakkola et al. 2019). AI systems are able to accomplish a given task as if they had the knowledge of an expert. These tasks can be visual perception, speech recognition, scene interpretation, object detection, decision-making, or translation. While performing all these tasks, problem-specific features should be obtained, and these features should be processed. This is the only thing that hasn't changed about AI since the past. In the past, by using simpler algorithms, an automatic decision-making process was carried out by using features such as edge information, frequency changes, plane differences (Crane 1979). In the following years, as the computing ability of computers increased, more sophisticated algorithms began to be used for feature extraction and classification. In the works of this period, certain features are usually selected based on the problem. Selected features are classified with a classifier. These methods, which require a lot of experience, were later named hand-crafted feature extraction (Majtner et al. 2016). AI methods have entered a new era after the emergence of deep learning. With the provision of a sufficient number of training examples, deep learning architectures automatically learn the features and classify these features automatically. Convolutional neural network (CNN) architectures, which are used primarily for image processing, appear in almost all new scientific researches today (Öztürk 2020). All these positive developments in the field of AI might help researchers today to fight COVID-19.

Considering the breakthrough results of CNN methods in the field of image processing, it is thought that CT and X-Ray images of COVID-19 will be diagnosed with very high performance. However, since CNN architectures have quite a lot of trainable parameters, they need as many training samples to train these parameters. For this reason, CNN is not the most suitable solution for datasets that do not contain enough samples. Data augmentation methods are preferred to solve this problem (Shorten and Khoshgoftaar 2019). The early datasets created for the analysis of CT and X-Ray images related to COVID-19 do not contain a sufficient number of training samples. When the number of samples in other classes is increased to increase the number of samples in these datasets, the number of COVID-19 class remains quite low. For this reason, it is challenging to see a successful CNN architecture in early COVID-19 classification and segmentation studies. Pereira et al. (Pereira et al. 2020) classify chest X-ray (CXR) for COVID-19 diagnosis using texture features such as local binary pattern (LBP), elongated quinary patterns (EQP), Local directional number (LDN), and binarized statistical image features (BSIF). Li et al. (Li et al. 2020) used hand-crafted image features on a chest CT dataset consisting of 78 patients in total. In this early study, they argued that some chest CT images could lead to misdiagnosis when used alone. Barstugan et al. (Barstugan et al. 2020) proposed a classification framework using a dataset consisting of a total of 150 CT images, 53 infected. They extracted Gray Level Co-occurrence Matrix (GLCM), Local Directional Patterns (LDP), Gray Level Run Length Matrix (GLRLM), Gray Level Size Zone Matrix (GLSZM), and Discrete Wavelet Transform (DWT) features and the features were classified by SVM. Shi et al. (Shi et al. 2020b) used location-specified features and feature selection for effective classification of CT images. Elaziz et al. (Damasevicius et al. 2020) have proposed a machine learning method for chest x-ray image analysis for COVID-19 detection. Their Fractional Multichannel Exponent Moments (FrMEMs) method is a hand-crafted feature extraction approach. Some researchers who want to take advantage of the power of CNN algorithms have used hand-crafted features and CNN features together. Altan and Karasu (Altan and Karasu 2020) presented a hybrid model consisting of curvelet transformation, chaotic salp swarm algorithm (CSSA) and deep learning to analyze X-ray images. Farid et al. (Farid et al. 2020) used hand-crafted features and deep features to analyze CT images.

As a result of the rapid sharing of CT and X-Ray images obtained in hospitals, the number of samples in COVID-19 datasets has started to increase rapidly. This has enabled the use of CNN architectures. Nour et al. (Nour et al. 2020) have optimized a linear CNN architecture with a Bayesian optimization model. Besides, they increased the number of positive samples with data augmentation. Ozkaya et al. (Ozkaya et al. 2020) performed a feature fusion using ResNet-50, GoogleNet, and VGG-16 deep architectures for effective COVID-19 detection using CT images. Their findings show that deep feature fusion can produce highly effective results. Wang et al. (Wang et al. 2020) have proposed a weakly-supervised framework for the rapid detection of COVID-19, which requires more samples for training. Apostolopoulos and Mpesiana (Apostolopoulos and Mpesiana 2020) carried out a study evaluating the transfer learning performance of CNN methods using a dataset consisting of a total of 1427 X-ray images encompassing 224 Covid-19 cases. Afshar et al. (Afshar et al. 2020) proposed a deep learning approach called COVID-CAPS for Covid-19 detection from X-Ray images. The capsule network method they recommend has produced remarkable results since it can work in small datasets. Jaiswal et al. (Jaiswal et al. 2020) used a pre-trained DenseNet201 architecture to get rid of the adverse effects of small datasets on CNN training. Brunese et al. (Brunese et al. 2020) propose and evaluate an approach based on transfer learning by exploiting the VGG-16 model.

When the CNN methods in the literature are examined, it is possible to find highly competent and high-performance methods. However, it is not efficient to use these methods with the current Covid-19 datasets in their original form because of datasets containing an insufficient number of samples and uneven distribution. Considering the workload of radiology experts, it is understood that it is not possible for now to create a sufficiently large labeled dataset. For this reason, many researchers avoided these powerful methods, while many researchers conducted studies on weakly-supervised methods or data augmentation. Although the results of the studies examined above are very inspiring, there is almost no study that can achieve satisfactory success in satisfactory times. For this reason, the main purpose of this study is to propose a powerful deep learning method for Covid-19 datasets containing an insufficient number of samples without encountering overfitting problems. Unlike many methods in the literature, we do not change the layer sequence of architectures or loss function of existing CNN structures. Because these architectures continue to have quite a lot of trainable parameters, and this still carries the overfitting problem. In this study, we use a convolutional support vector machine (CSVM) network (Bazi and Melgani 2018). To the best of our knowledge, this is the first study in which uses CSVM to classify CT images for Covid-19 detection. CSVM network consists of blending several convolutions and pooling layers ending with the SVM layer. SVMs are quite decisive for the production of filter banks. The learning process, on the other hand, takes place as forward supervised learning, in contrast to the standard backpropagation approach. In this way, a very high learning performance occurs despite limited training data. The main contributions of this study are:

- Since the proposed method is designed for Covid-19 datasets containing a small number of samples, it has higher classification performance than other state-of-the-art methods in the literature.
- Compared to the familiar CSVM architecture, the number of parameters is reduced, and it can be trained using fewer training samples.
- Test time is faster than other deep architectures in the literature.

The rest of this paper is organized as follows. Section 2 provides a methodological background and proposed method details. Dataset and experimental details are presented in Section 3. Also, comparison tables and results are given in this section. Section 4 presents the discussion and section 5 conclusion.

## 2. METHODS

In this section, the background information of the proposed architecture methodology, proposed technique and application details, model architecture, and information about the parameters will be presented.

### 2.1. Background

In order for an AI algorithm to produce decisions similar to the decision-making process of an expert, it must perform a similar evaluation process. For image processing problems, the decision-making cycle of an expert usually consists of evaluating the input taken through his eyes using his brain. Visually obtained information consists of many parts from low-level to high-level. While image processing algorithms used only low-level algorithms in the early period, the situation is changing today. CNN architectures, which are very effective for image processing problems, can automatically learn low-level, mid-level, and high-level features (Öztürk and Özkaya 2020). These learned features are kept in convolution kernels. Convolution kernels or in other words convolution layers are one of the most important layers of CNN architecture. The classical convolution process is applied by sliding each convolution kernel of the convolution layer on the image. Since each kernel is shifted over the entire image, there is a severe decrease in the number of parameters. Parameters in convolution kernels are updated with the backpropagation process and learn the features of the problem. Although the number of trainable parameters is reduced compared to a fully connected layer thanks to the parameter sharing in the convolution layer, the number of parameters in the CNN architecture is still quite high. (Fig. 2.1) shows how the process occurs in convolution neural networks:

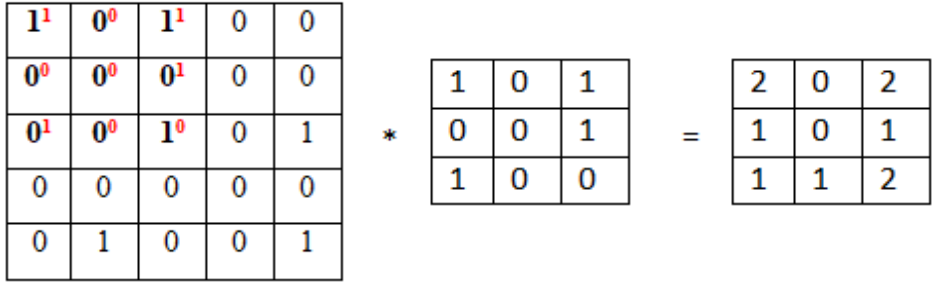

**Fig. 2.1** Convolution operation

The size of the feature map obtained as output is calculated as in (Eq. 2.1):

$$Nout = \frac{Nin + 2P - K}{S} + 1 \qquad (2.1)$$

'$N_{in}$' input is the size of the input image, '$P$' is the number of zero layer to be added around the image, '$K$' is the size of the filter, '$S$' is how many steps the filter moves at each time, and '$N_{out}$' is the output size.

The stride parameter '$S$' is the parameter expressing how many units the convolution filter will shift in each stage. The value determines the size of the feature map. Figure 2.2 shows the feature matrix representations that occur for the cases where s value is 1 and 2.

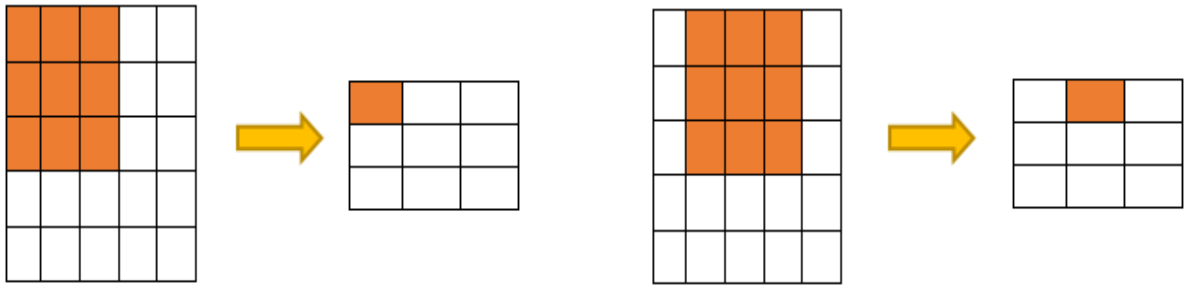

a

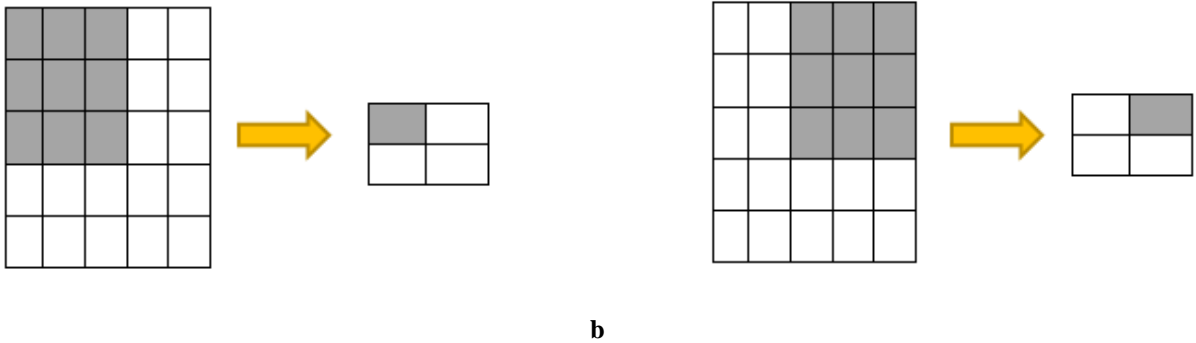

**Fig. 2.2** Feature maps with 'S' value **a** S=1, **b** S=2

As can be seen from (Fig. 2.2), after the convolution process, the size of the feature map is lower than the size of the original image. Zero addition is made around the original image where size reduction is desired, as seen in (Fig. 2.3).

**Fig. 2.3** Zero-padding

Another important layer, the pooling layer, where important parameters are preserved, and insignificant parameters are eliminated, overcomes this problem. In the pooling process, the square window size is determined first. This window is scrolled over the image like convolution kernels. The pooling process is applied to the part under the pooling window. It is expected to be selected in an order in the encoding size as the size of the size of the pool becomes smaller in size. In the literature, max-pooling, sum-pooling, and average-pooling are generally used. In the max-pooling process, the maximum value is selected among the pixels under the pooling window. In average-pooling, the arithmetic average of the pixels under the pooling window is used. In the sum-pooling process, the total value of the pixels under the pooling window is calculated, and this value is transmitted. 2x2 pooling process is shown in (Fig. 2.4).

7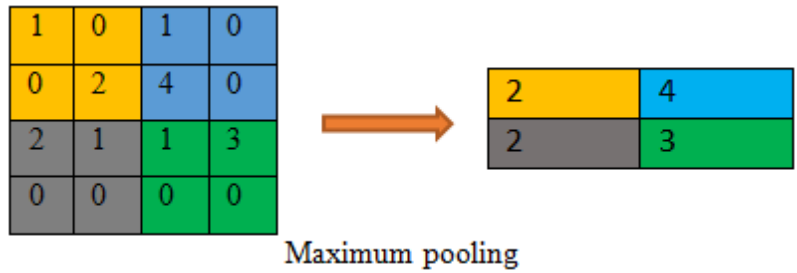

Maximum pooling

a

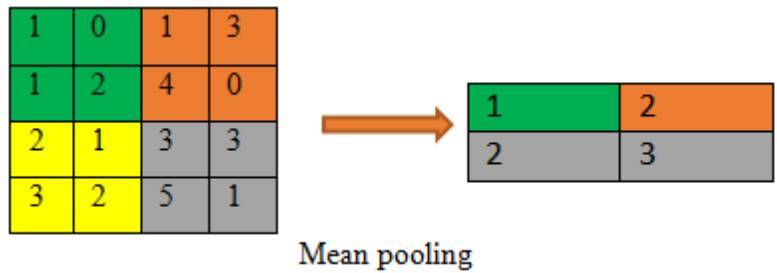

Mean pooling

b

**Fig. 2.4** Maximum and mean pooling representation, **a** maximum pooling, **b** mean pooling

Another important layer of CNN architectures is the activation function layer. Activation functions are functions that bring incoming input values to a certain range or allow some of the input values to be taken and some of them cut. The most frequently encountered activation functions in the literature are sigmoid (Eq. 2.2), tanh (Eq. 2.3) and ReLU (Eq 2.4) activation functions.

$$s = \frac{1}{1+e^{-z}} \tag{2.2}$$

$$t = \frac{1-e^{-z}}{1+e^{-z}} \tag{2.3}$$

$$r = \max(0, z) \tag{2.4}$$

where 'z' represent the input value and the output value of activation functions 's', 't' and 'r' sigmoid, tanh and relu respectively. ReLU activation function process is shown in (Fig. 2.5) (Agarap 2018).

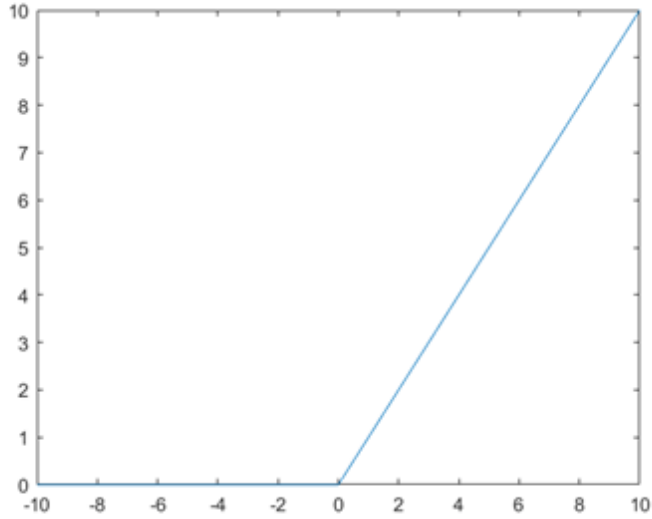

**Fig. 2.5** The ReLU activation function

Today, the ReLU activation function is used as a standard in almost all CNN architectures. This layer changes the parameters that are negative in value to zero. The output of a CNN architecture consisting of only these three essential layers can be calculated as in (Eq. 2.5).

$$f(o) = \beta_{nxn}\left(\sigma(w \times I + b)\right) \tag{2.5}$$

where $I$ represents the input image, $o$ represents the output of CNN, $f$ is CNN operation, $\beta$ represents max-pooling layer, $n$ represents the kernel size of the pooling layer, $\sigma$ represents the ReLU operation, $w$ is the weights of the CNN layer, b is bias value, and $x$ represents convolution operation. Features automatically obtained with the help of these basic layers are classified with a fully connected layer (FCL). FCL is a multi-layered perceptron (MLP) structure. In FCL output, the probability distribution is realized with a softmax layer. In addition to these basic layers, various layers, such as the concatenate layer and normalization layer, are recommended to increase performance day by day. In addition to linear CNN architectures, various new architectures such as residual architectures and parallel architectures enter the literature.

One of the main reasons underlying CNN remarkable performance is huge datasets. If CNN architectures are not trained with datasets containing a sufficient number of labeled samples, either they cannot learn the problem or overfitting may occur. This data dependence is one of the most serious problems of deep learning. The data labeling process is very laborious and lengthy. In addition, it is impossible to find enough labeled data for some problems such as medical data, sudden emergencies. The transfer learning approach is generally a very effective tool to solve this problem. Briefly working on transfer learning is as follows: CNN architecture is trained with a labeled huge dataset, and CNN learns low, mid, and high-level features. Then, the information obtained is transferred from the source domain to the target domain. At this stage, the assumption that training data and test data must be independent and uniformly distributed random variables is relaxed (Tan et al. 2018). Let's have a domain represented by $D=\{x,$

*P(x)}*, where *x* represents feature space and *P(x)* represents a probability distribution. Let's define a task with *T={y, f(x)}*, where *y* represents label space and *f(x)* is target prediction function. When $T_t$ is given a learning task based on $D_t$, $D_s$ for the learning task $T_s$ can be used for help in this task. The transfer learning approach aims to increase the predictive power of the $f_t(.)$ function here and to do the information transfer task in the best way.

CNN architectures classify data using the MLP structure in its last layer. SVM has come to the fore lately because MLP needs more training samples than SVM and is more vulnerable to some problems. SVM is less sensitive to the overfitting problem due to margin maximization.

**2.2. Proposed Method**

Support Vector Machine (SVM) is one of the most popular supervised classifiers in the literature. It is a method that solves the overfitting problem using the maximum margin (distance) principle. SVM algorithm is trained with manual or automatic extracted features to solve various recognition problems. Convolutional Support Vector Machine (CSVM) usually works together with a CNN feature extractor to fulfill the classification task (Bazi and Melgani 2018)**.** In this algorithm, convolution filters are used to obtain feature maps. Compared to standard CNN structures, the CSVM algorithm is a deep learning technique based on the SVM method. It does not use a backpropagation algorithm to update weights and biases.

CSVM network is based on various alternative convolution and pooling layers, followed by a classification layer. Each convolution layer uses a series of linear SVM filter banks to generate a new feature map. For the first convolutional layer, SVM filters process the original input images. SVM weights of each convolution layer are calculated in a directly supervised manner by working on images representing the objects of interest. The generated feature maps by convolution layers are then processed in a nonlinear activation function such as ReLU. The pooling layer in CSVM works similarly as in CNN structures. It takes small rectangular patches from the convolutional layer and produces a single output from each patch. Finally, higher-level feature vectors are given input into a linear SVM classifier to fulfill the classification task.

The convolutional SVM layer is the main structure of the CSVM network. This layer uses linear SVM weights as convolutional filters to generate feature maps. These filters learn weights by using a feed-forward learning method, unlike traditional CNN's weights by backpropagation. In the dataset, positive images contain the object of interest, while negative images represent the background. Firstly, an image patch set with three channels $I_i$, $h \times h \times 3$ is extracted from each image. After all images are processed, the data set defined as $Tr^{(1)}$ is generated. A set of SVM filters (SVMs) are learned from randomly selected training data from different sub-training sets allocated. The weight vector $w \in R^d$ and bias values $b \in R$ of each SVM filter are determined by optimizing objective function as in (Eq. 2.6).

$$\min_{w,b} w^T w + C \sum_{i=1}^{l} \xi(w,b;x_i,y_i) \tag{2.6}$$

$C$ is assigned as the penalty variable. The square hinge cost function $max(1-y_i(w^T x^i+b),o)^2$ was used with reference to L1-SVM and L2-SVM. To put it more simply, bias values have been omitted from the formula and SVM filters can be trained in this form of the formula. Subsequently, all the weights of the convolution layer can be grouped into a four-dimensional filter stack. In the creation of the convolution feature map, each training image enters the convolution process with SVM filters. Convolution process on the image is mathematically shown in (Eq. 2.7).

$$h_{ki}^{(1)} = f(I_i * w_k^{(1)}), \quad k=1,......,n^{(1)} \tag{2.7}$$

$h_{ki}$ obtained $k$. shows the feature map as in Eq. 3. The * operator indicates the convolution process. $f$ is the activation function of ReLU

(Fig. 2.6) contains CSVM models with different layer numbers. The input image is in gray level and 128×128 dimension size. The first block has 40 SVMs in the first layer and has convolution windows of 7×7 dimensions. The stride amount of these filters is 2 × 2. Then, there is the ReLU layer as the activation function. There is a pooling layer to obtain higher-level features. In the maximum pooling level, 3×3 window size and 2×2 stride amount are processed. The obtained features are trained and tested with the linear SVM classifier. Also, these feature maps are then transferred to the latter block with a different number of 128 SVMs. In this block, convolution filters are 3 × 3 in size and have a 1×1 stride. There are ReLU layers and pooling layers with the same properties as in the previous block. Linear SVM algorithm is used as a classifier. There are 256 SVMs in the last block characterized by 1x1 convolution windows and 1x1 stride size. The obtained feature maps are passed through the ReLU activation function, and feature selection is made with the pooling layer. The training and testing of the linear SVM classifier are carried out with the last high level of features.

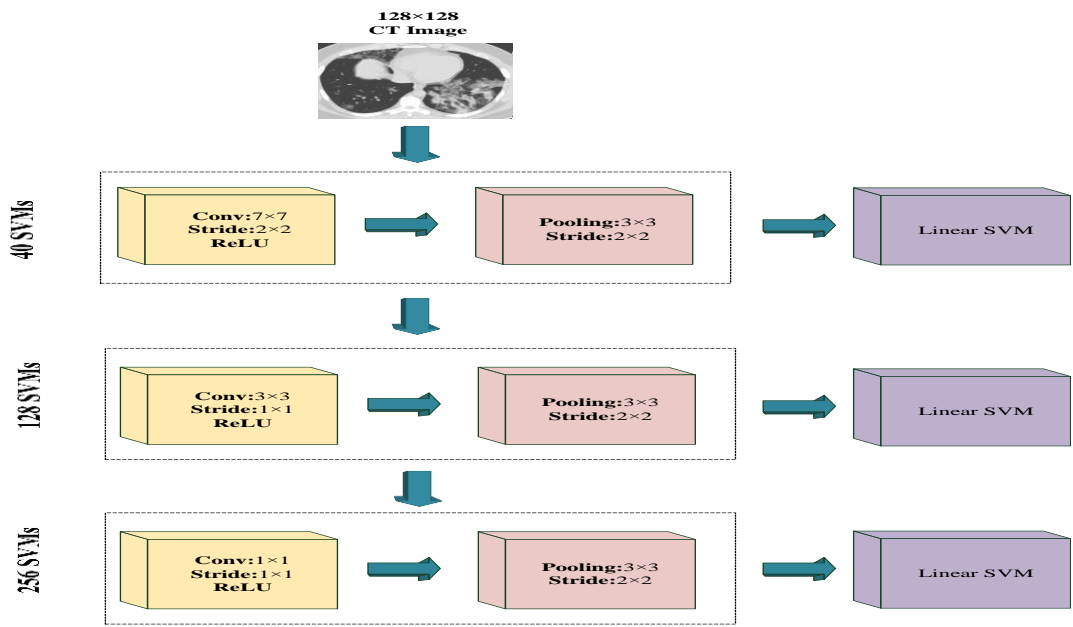

**Fig. 2.6** Proposed CSVM Models

## 3. RESULTS

### 3.1. Dataset

In this study, the SARS-CoV-2 CT scan dataset (www.kaggle.com/plameneduardo/sarscov2-ctscan-dataset), which includes a binary classification problem, is used. The data set includes 2492 CT-scans images in total. While 1262 images are positive (COVID-19), 1230 images are negative (non-COVID). The images in the data set are divided into 75% and 25% for training and testing for pre-train CNN structures and CSVM models. Data augmentation methods are not applied to the data set. Some positive (COVID-19) and negative (non-COVID) image samples are reported in (Fig. 3.1).

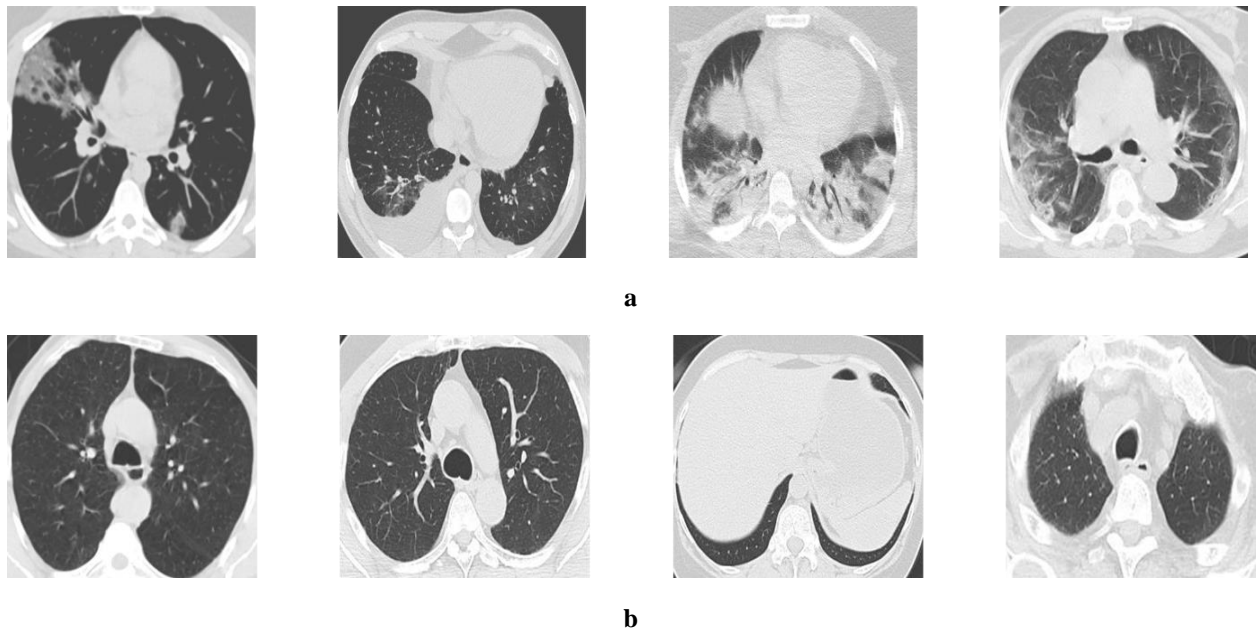

**Fig. 3.1** Samples of CT Images **a** Positive (COVID-19), **b** Negative (non-COVID)

### 3.2. Evaluation Metrics

Seven different classification metrics are used to evaluate pre-trained CNN models and CSVM. These metrics are Accuracy (ACC), Sensitivity (SEN), Specificity (SPE), Precision (PRE), F1-Score, Matthews Correlation Coefficient (MCC), and Kappa as in (Eqs. 3.1-3.7). At the same time, training time is another evaluation criterion. Classification metrics are computed with True Positive (TP), True Negative (TN), False Positive (FP) and False Negative (FN).

$$Accuracy = (TP + TN) / (TP + FN + TN + FP) \tag{3.1}$$

$$Sensitivity = TP / (TP + FN) \tag{3.2}$$

$$Specificity = TN / (TN + FP) \tag{3.3}$$

$$Precision = TP / (TP + FP) \tag{3.4}$$

$$F1 - Score = (2 \times TP) / (2 \times TP + FN + FP) \tag{3.5}$$

$$MCC = \frac{TP \times TN - FP \times FN}{\sqrt{(TP + FP)(TP + FN)(TN + FP)(TN + FN)}} \tag{3.6}$$

$$Kappa = (total\ accuracy - random\ accuracy) / (1 - random\ accuracy) \tag{3.7}$$

**3.3. Findings and Discussion**

The experiments in the study are carried out on an Intel Core i7-7700 HQ CPU at 2.8 GHz, 16 GB RAM, and NVIDIA GTX 1080 GPU. Matlab 2020a is used as a simulation program. Pre-trained CNN networks have been trained up to 30 Epochs. The mini-batch size is set at 10. Each epoch is completed in 186 iterations. The initial learning rate is 0.1, and a drop factor of 0.1 is applied to each ten epoch. Stochastic gradient descent with momentum has been chosen as the optimization method. ResNet-50 model has achieved the highest accuracy in pre-train CNN networks. The training and test graphics of this model are given together in (Fig. 3.2) In the accuracy graph, the blue curves show the training data accuracy, while the black curves represent the test data accuracy. In the loss graph, red curves represent the loss of training, black curves represent loss for test data.

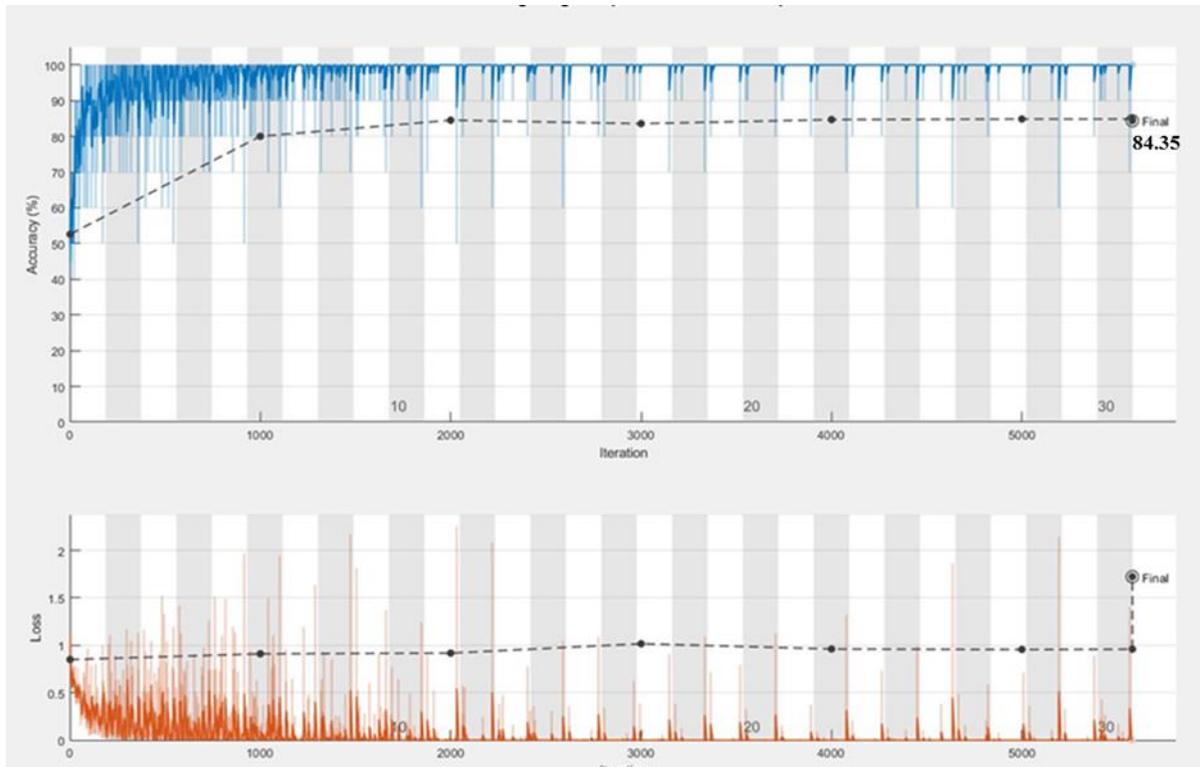
**Fig. 3.2** ResNet-50 Training and Testing Process

A confusion matrix is needed to measure classification performance. In (Fig. 3.3), there are confusion matrices for pre-train CNN models in test data. The test data includes 313 COVID and 307 non-COVID images. Pre-train CNN networks training is carried out using the transfer learning method. In the VGG-16 model, the number of TP is 213, and FN is 100. In this pre-train CNN model, only 1 out of 307 non-COVID data was misclassified. Also, 68.1% accuracy was obtained in the COVID class and 99.7% accuracy in the non-COVID class. The error rate is 31.9% for COVID and 0.3% for non-COVID class. When the results obtained by the ResNet-50 model are examined, 216 images were classified correctly in the COVID class, and 97 images were classified incorrectly. The obtained accuracy percentage for the COVID class is 69.8%, and the error percentage is 30.2%. All images in the non-COVID class have been classified correctly, and 100% class accuracy has been achieved. GoogleNet, another pre-trained CNN model, has 206 TP and 107 FN in the COVID class. When the non-COVID class is evaluated, there are 5 FP and 306 TN. 65.8% for the COVID class, and 98.4% for the non-COVID class was obtained. Error rates are 34.2% and 1.6%, respectively, in terms of classes. Another pre-train CNN network in the study is DenseNet-201. For the metric evaluation of this network, 196 TP, 117 FN, 1 FP, and 306 TN values were reached. When the class accuracy percentages are examined, 62.6% for the COVID class, and 99.7% for the non-COVID class were obtained. Error percentages are 37.4% and 0.3%, respectively.

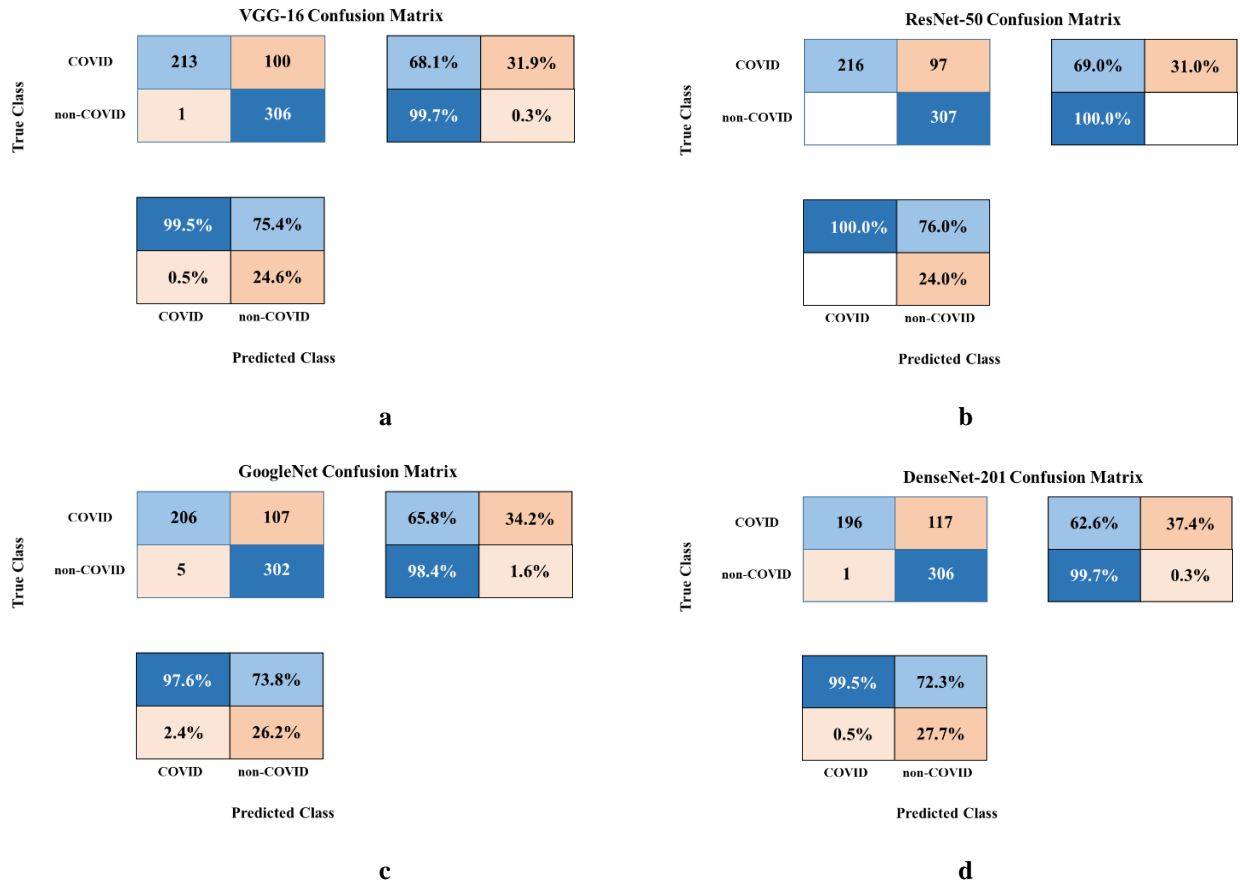

**Fig. 3.3** Confusion Matrix of pre-trained CNNs **a** VGG-16, **b** ResNet-50, **c** GoogleNet, **d** DenseNet-201

Confusion matrices of proposed CSVM models for the test data are shown in (Fig. 3.4). In addition to the pre-trained CNN models trained using transfer learning, the confusion matrices of the proposed CSVM models are also calculated. When Fig. 5 is examined, it is seen that the accuracy rate obtained for the COVID class is not sufficient. It is seen that this problem has been largely overcome with proposed CSVM models. In the confusion matrix for the CSVM (7×7) model, 284 TP, 29 FN, 40 FP, and 267 TN values were obtained. When the class accuracy percentages are considered, it has been reached 90.7% for the COVID class and 87.0% for the non-COVID class. Error rates are 9.3% and 13.0%, respectively. The CSVM (7×7, 3×3) model, on the other hand, increased the performance slightly more than the other CSVM model. It obtained 289 TP, 24 FN, 23 FP, and 284 TN values. COVID and non-COVID class accuracy are 92.3% and 92.5%, respectively. Their class error percentages are 7.7% and 7.5%. CSVM (7×7, 3×3, 1×1) showed the highest performance among CSVM models. Thanks to this model, 288 TP, 25 FN, 12 FP, and 295 TN values were reached. When the COVID and non-COVID class accuracy rates were examined, 92.0% and 96.1% were obtained. Error rates are at levels of 8.0% and 3.9%.

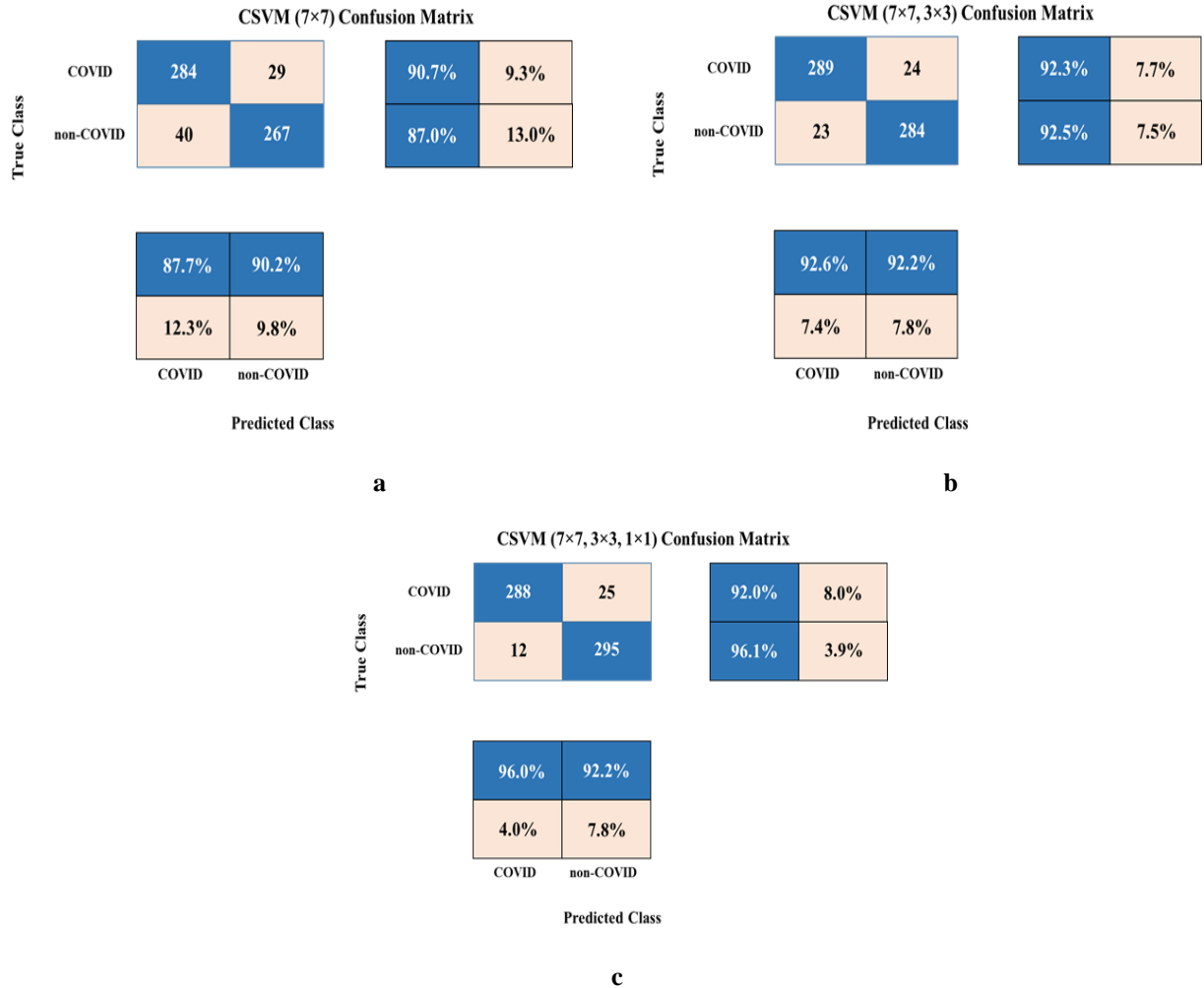

**Fig. 3.4** Confusion Matrix of CSVM Models **a** (7×7), **b** (7×7, 3×3), **c** (7×7, 3×3, 1×1)

## 4. DISCUSSION

Classification performances of the pre-trained CNN models and CSVM models are given in Table. The CSVM (7×7, 3×3, 1×1) model achieved the highest performance with 94.03% ACC, 96.09% SEN, 94.10% F1-Score, 88.15% MCC and 88.07% Kappa. The ResNet-50 model has the highest metric values of 100% SPE and 100% PRE. The model with the longest training period is the DenseNet-201 model with 636 min 34 sec. The shortest training time belongs to the 7min 46sec CSVM (7×7) model.

(Fig. 4.1) includes Receiver Operating Characteristic (ROC) curves and Area Under Curve (AUC) values of pre-trained CNN models and CSVM models. When AUC metric values are evaluated, GoogleNet has the lowest AUC value with 0.8944. DenseNet-201 has the highest AUC metric in pre-trained CNN models with 0.9356. The AUC metric values of CSVM models are higher than pre-trained CNN models. The 0.9866 AUC value obtained using the CSVM (7×7, 3×3, 1×1) model has the highest performance.

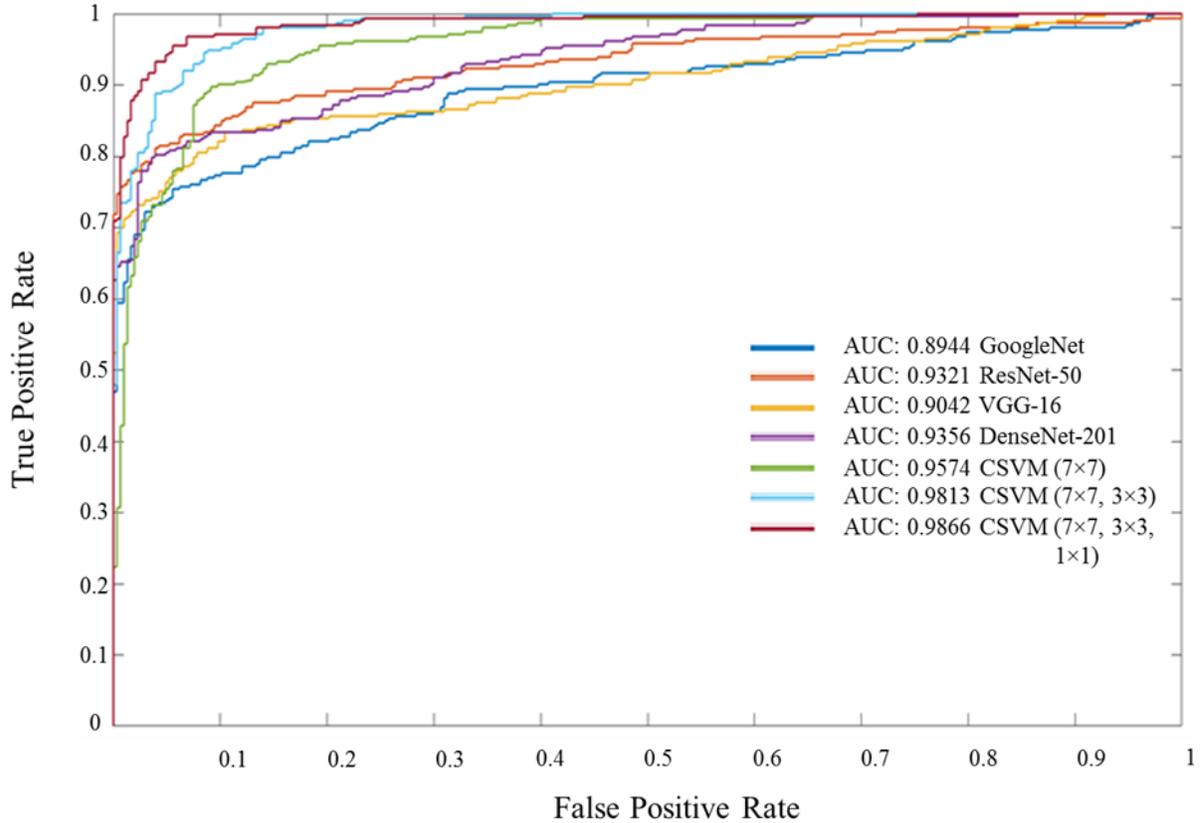

**Fig. 4.1** ROC Curves for Deep learning Methods

The impact of COVID-19, which has become a global epidemic, is assumed to show in the future. Besides, it is known that the days when the effect we call the second wave will increase are approaching. Thanks to this study, a COVID-19 diagnosis system has been proposed to carry out clinical studies quickly and automatically. The proposed model is based on CSVM architecture and shows higher performance than other pre-trained CNN networks. At the same time, the training time is very low compared to CNN models. CSVM models were trained as scratch models, contain less number of parameters than pre-trained CNN models. The CSVM algorithm, which can provide automatic feature extraction, can analyze CT images with high precision. In this study, the CSVM model provides high performance and low training time. CSVM (7×7, 3×3, 1×1) model achieved 94.03% ACC, 96.09% SEN, 92.01% SPE, 92.19% PRE, 94.10% F1-Score, 88.15% MCC and 88.07% Kappa metric values. It is thought that it can support the processes of instant clinical evaluation and supporting expert opinions by means of the CSVM model. In addition, by comparing with other methods in the literature, we have provided the classification and training time results in (Table 4.1).

Table 4.1. Classification and Training Time Results

| Methods | Evaluation Metrics (%) | | | | | | | Training Time |
|---|---|---|---|---|---|---|---|---|
| | ACC | SEN | SPE | PRE | F1-score | MCC | Kappa | |
| VGG-16 | 83.71 | 68.05 | 99.67 | 99.53 | 80.83 | 71.22 | 67.52 | 55 min 37 sec |
| ResNet-50 | 84.35 | 69.01 | **100** | **100** | 81.66 | 72.42 | 68.80 | 81 min 6 sec |
| GoogleNet | 81.94 | 65.81 | 98.37 | 97.63 | 78.63 | 67.73 | 63.98 | 47 min 9 sec |
| DenseNet-201 | 80.97 | 62.62 | 99.67 | 99.49 | 76.86 | 66.89 | 62.07 | 636 min 34 sec |
| CSVM (7×7) | 88.87 | 86.97 | 90.73 | 90.20 | 88.56 | 77.78 | 77.73 | **7 min 46 sec** |
| CSVM (7×7, 3×3) | 92.42 | 92.51 | 92.33 | 92.21 | 92.36 | 84.84 | 84.84 | 15 min 11 sec |
| CSVM (7×7, 3×3, 1×1) | **94.03** | **96.09** | 92.01 | 92.19 | **94.10** | **88.15** | **88.07** | 25 min 36 sec |

## 5. CONCLUSION

In this study, . The data set includes 2492 CT-scans images in total. While 1262 images are positive (COVID-19), 1230 images are negative (non-COVID). The images in the data set are divided into 75% and 25% for training and testing for pre-traineed CNN structures and CSVM models. CSVM model provides high performance and low training time. CSVM (7×7, 3×3, 1×1) model achieved 94.03% ACC. Looking at the performance of the developed CSVM model, it can be evaluated by experts and ready to be tested with a larger dataset CSVM is planned to be trained with larger datasets to achieve the level of success that can assist physicians in the diagnosis of COVID-19 disease. A training process with large data sets is very important in determining the validity and reliability of the system. This developed system can be used to address the shortage of radiologists due to the increasing number of cases in countries affected by COVID-19. Also, such models can be used to diagnose other chest-related diseases. We plan to make our model more accurate and robust by obtaining more of these types from our hospitals.

**Compliance with Ethical Standards**

Conflict of interest: The authors declare that they have no conflicts of interest.

Human and animal rights: The paper does not contain any studies with human participants or animals performed by any of the authors.

**Main Points:**

- The proposed method offers a very fast and high-performance framework to actively combat COVID-19. In this way, it reduces the workload of medical experts.
- To the best of our knowledge, the CSVM method is actively used for the first time in this study to detect COVID-19.
- The proposed framework produces better performance metrics than other state-of-the-art deep learning techniques.
- For datasets containing a small number of samples and a small variety of samples (eg nearly all COVID-19 datasets), the recommended method is ideal.

**List the figure legends:**

Fig. 2.1 Convolution operation

Fig. 2.2 (a) Feature maps with 'S' value 1

Fig. 2.2 (b) Feature maps with 'S' value 2

Fig. 2.3 Zero-padding

Fig. 2.4 (a) Maximum pooling, b mean pooling

Fig. 2.4 (a) Mean pooling

Fig. 2.5 The ReLU activation function

Fig. 2.6 Proposed CSVM Models

Fig. 3.1 (a) Samples of Positive CT Images (COVID-19)

Fig. 3.1 (b) Samples of Negative CT Images (non-COVID)

Fig. 3.2 ResNet-50 Training and Testing Process

Fig. 3.3 (a)  Confusion Matrix of VGG-16

Fig. 3.3 (b)  Confusion Matrix of ResNet-50

Fig. 3.3 (c)  Confusion Matrix of GoogleNet

Fig. 3.3 (d)  Confusion Matrix of DenseNet-201

Fig. 3.4 (a)  Confusion Matrix of (7×7) CSVM Models

Fig. 3.4 (b)  Confusion Matrix of (7×7) CSVM Models

Fig. 3.4 (c)  Confusion Matrix of (7×7) CSVM Models

Fig. 4.1 ROC Curves for Deep learning Methods